\documentclass[a4paper,english]{article}
\usepackage[T1]{fontenc}
\usepackage[latin1]{inputenc}
\usepackage{amsmath}
\usepackage{graphicx}
\usepackage{amssymb}

\makeatletter

\providecommand{\tabularnewline}{\\}

 \newcommand{\lyxaddress}[1]{
   \par {\raggedright #1 
   \vspace{1.4em}
   \noindent\par}
 }

\usepackage{babel}
\makeatother
\begin{document}

\title{\textbf{\Large Gap Equations and Electroweak Symmetry Breaking}}

\author{G. Cynolter and E. Lendvai}

\date{$\phantom{.}$}

\maketitle

\lyxaddress{\textit{Theoretical Physics Research Group of Hungarian Academy of
Sciences, Eötvös University, Budapest, 1117 Pázmány Péter sétány 1/A,
Hungary}\\
 \textit{}}

\begin{abstract}
Recently a new dynamical symmetry breaking model of electroweak interactions
was proposed based on interacting fermions. Two fermions of different
$SU_{L}(2)$ representations form a symmetry breaking condensate and
generate the lepton and quark masses. The weak gauge bosons get their
usual standard model masses from a gauge invariant Lagrangian of a
composite doublet scalar field. The new fermion fields become massive
by condensation. In this note the gap equations are given in the linearized
(mean field) approximation and the conditions for symmetry breaking
and mass generation are presented. Perturbative unitarity constrains
the self-couplings and the masses of the new fermions, a raw spectrum
is given.
\end{abstract}
With the advent of the LHC the problem of electroweak symmetry breaking
becomes more and more important. The LHC experiments are expected
to shed light on the dynamics of the symmetry breaking. As there is
no direct evidence for elementary scalar (Higgs) particles, alternative,
dynamical mechanisms have also been investigated, like technicolour,
top condensate or topcolour models \cite{dynsym,top,vcm04}. These
models were reborn in extra dimensional scenarios like higgsless models
\cite{higgsl}. Recently a new dynamical symmetry breaking model was
proposed based on new fermions of different representations of the
weak gauge group \cite{fcm}, in the model four-fermion interactions
generate the (symmetry breaking) condensates. The condensates generate
masses for the new fermions and an auxiliary composite scalar field
is responsible for the weak gauge boson masses. The model is non-renormalizable
and regulated by a four dimensional cutoff, we present the model after
the introduction in more detail. In this work we extend the model
to include more general condensates and investigate the non-conventional
structure of the gap equations in the modified model. We were able
to analyse the coupled gap equations and give the condition of finding
a symmetry breaking solution and fulfilling the constraints of perturbative
unitarity. The possible masses for various coupling constants are
presented and it is shown that the model can generate fermions with
a mass of few hundred GeV for cutoffs in the TeV range.

To start with we summarize the fermion condensate model of electroweak
interactions \cite{fcm}. In the model to be outlined the Higgs sector
is replaced with new fermions with non-renormalizable four-fermion
interactions. Under $SU_{L}(2)\times U_{Y}(1)$ the new fermions are
a neutral singlet $\Psi_{S}$, and a weak doublet $\Psi_{D}=\left(\begin{array}{c}
\Psi_{D}^{+}\\
\Psi_{D}^{0}\end{array}\right)$ with hypercharge 1. $\Psi_{D}^{+}\left(\Psi_{D}^{0}\right)$ is a
positively charged (neutral) field. The new fermions have effective
four-fermion interactions, valid up to some physical cutoff, the ultraviolet
completion of the model is not specified.

The new Lagrangian with gauge invariant kinetic terms and invariant
4-fermion interactions of the new fermions is $L_{\Psi}$,\begin{eqnarray}
L_{\Psi} & = & \phantom+i\overline{\Psi}_{D}D_{\mu}\gamma^{\mu}\Psi_{D}+i\overline{\Psi}_{S}\partial_{\mu}\gamma^{\mu}\Psi_{S}-m_{0D}\overline{\Psi}_{D}\Psi_{D}-m_{0S}\overline{\Psi}_{S}\Psi_{S}+\nonumber \\
 &  & +\lambda_{1}\left(\overline{\Psi}_{D}\Psi_{D}\right)^{2}+\lambda_{2}\left(\overline{\Psi}_{S}\Psi_{S}\right)^{2}+2\lambda_{3}\left(\overline{\Psi}_{D}\Psi_{D}\right)\left(\overline{\Psi}_{S}\Psi_{S}\right),\label{eq:4fermion}\end{eqnarray}
$m_{0D},m_{0S}$ are bare masses and $D_{\mu}$ is the covariant derivative
\begin{equation}
D_{\mu}=\partial_{\mu}-i\frac{g}{2}\underline{\tau}\,\underline{A}_{\mu}-i\frac{g'}{2}B_{\mu},\label{eq:covariantd}\end{equation}
where $\underline{A}_{\mu,}B_{\mu}$ and $g,\; g'$ are the usual
weak gauge boson fields and couplings, respectively. Additional four-fermion
couplings are possible but those will not basically change the symmetry
breaking and mass generation. We will show in what follows that for
couplings $\lambda_{i}$ exceeding the critical value the four-fermion
interactions of (\ref{eq:4fermion}) generate condensates 

\begin{eqnarray}
\left\langle \overline{\Psi}_{D\alpha}^{0}\Psi_{D\beta}^{0}\right\rangle _{0} & = & a_{1}\delta_{\alpha\beta},\label{eq:condD}\\
\left\langle \overline{\Psi}_{D\alpha}^{+}\Psi_{D\beta}^{+}\right\rangle _{0} & = & a_{+}\delta_{\alpha\beta},\label{eq:condP}\\
\left\langle \overline{\Psi}_{S\alpha}\Psi_{S\beta}\right\rangle _{0} & = & a_{2}\delta_{\alpha\beta},\label{eq:condS}\\
\left\langle \overline{\Psi}_{S}\Psi_{D}\right\rangle _{0}=\left\langle \left(\begin{array}{c}
\overline{\Psi}_{S}\Psi_{D}^{+}\\
\overline{\Psi}_{S}\Psi_{D}^{0}\end{array}\right)\right\rangle _{0} & \neq & 0\label{eq:conddoublet}\end{eqnarray}
 The formation of the charged condensate (\ref{eq:condP}) is more
general then in \cite{fcm}, but it does not change the original arguments
about symmetry breaking. The non-diagonal condensate in (\ref{eq:conddoublet})
spontaneously breaks $SU_{L}(2)\times U_{Y}(1)$ to $U_{em}(1)$ of
electromagnetism. With the gauge transformations of $\Psi_{D}$ the
condensate (\ref{eq:conddoublet}) can always be transformed into
a real lower component,

\begin{equation}
\left\langle \overline{\Psi}_{S\alpha}\Psi_{D\beta}^{0}\right\rangle _{0}=a_{3}\delta_{\alpha\beta},\quad\left\langle \overline{\Psi}{}_{S\alpha}\Psi_{D\beta}^{+}\right\rangle _{0}=0,\label{eq:mixed cond}\end{equation}
where $a_{3}$ is real. The composite operator $\overline{\Psi}_{S}\Psi_{D}$
resembles the standard scalar doublet. Assuming invariant four-fermion
interactions for the new and known fermions,\begin{equation}
L_{f}=g_{f}\left(\overline{\Psi}_{L}^{f}\Psi_{R}^{f}\right)\left(\overline{\Psi}_{S}\Psi_{D}\right)+g_{f}\left(\overline{\Psi}_{R}^{f}\Psi_{L}^{f}\right)\left(\overline{\Psi}_{D}\Psi_{S}\right),\label{eq:Yukawa}\end{equation}
the condensate (\ref{eq:mixed cond}) generates masses to the standard
femions. In the linearized, mean field approximation the electron
mass, for example, is\begin{equation}
m_{e}=-4g_{e}a_{3}.\label{eq:melectron}\end{equation}
Up type quark masses can be generated via the charge conjugate field
$\widetilde{\Psi}_{D}=i\tau_{2}\left(\Psi_{D}\right)^{\dagger}.$ 

The masses of the weak gauge bosons arise from the effective interactions
of the auxiliary composite $Y=1$ scalar doublet,\begin{equation}
\Phi=\left(\begin{array}{c}
\Phi^{+}\\
\Phi^{0}\end{array}\right)=\overline{\Psi}_{S}\Psi_{D}.\label{eq:scalardef}\end{equation}
$\Phi$ develops a gauge invariant kinetic term in the low energy
effective description 

\begin{equation}
L_{H}=h\left(D_{\mu}\Phi\right)^{\dagger}\left(D^{\mu}\Phi\right),\label{eq:L phi}\end{equation}
 where $D_{\mu}$ is the usual covariant derivative (\ref{eq:covariantd}).

The coupling constant $h$ sets the dimension of $L_{H}$, $[h]=-4$
in mass dimension, we assume $h>0$. (\ref{eq:L phi}) is a non-renormalizable
Lagrangian and it provides the weak gauge boson masses and some of
the interactions of the new fermions with the standard gauge bosons. 

The terms with $\Phi^{0}$ in $L_{H}$ can be written as\begin{eqnarray}
h^{-1}L_{H} & =\phantom+ & \frac{g^{2}}{2}W_{\mu}^{-}W^{+\mu}\Phi^{0\dagger}\Phi^{0}+\frac{g^{2}}{4\cdot\cos^{2}\theta_{W}}Z_{\mu}Z^{\mu}\Phi^{0\dagger}\Phi^{0}+\label{eq:hLH}\\
 &  & +\left[\partial^{\mu}\Phi^{0\dagger}\partial_{\mu}\Phi^{0}-\frac{i}{2}\frac{g}{\cos\theta_{W}}\left(\partial^{\mu}\Phi^{0\dagger}\right)\Phi^{0}Z_{\mu}+\frac{i}{2}\frac{g}{\cos\theta_{W}}\Phi^{0\dagger}Z_{\mu}\left(\partial^{\mu}\Phi^{0}\right)\right]\nonumber \end{eqnarray}
in terms of the usual vector boson fields.

In the linearized approximation in (\ref{eq:hLH}) we put

\begin{equation}
h\,\Phi^{0\dagger}\Phi^{0}\rightarrow h\left\langle \Phi^{0\dagger}\Phi^{0}\right\rangle _{0}=h\left(16a_{3}^{2}-4a_{1}a_{2}\right)=\frac{v^{2}}{2},\label{eq:phi vev}\end{equation}
leading to the standard masses\begin{equation}
m_{W}=\frac{gv}{2},\qquad m_{Z}=\frac{gv}{2\cos\theta_{W}}.\label{eq:standardm}\end{equation}
$v^{2}$ is, as usual, $\left(\sqrt{2}G_{F}\right)^{-1}$; $v=254$
GeV . The tree masses naturally fulfill the important relation $\rho_{\mathrm{tree}}=1$. 

Once the condensates (\ref{eq:condD}-\ref{eq:conddoublet}) are formed,
dynamical mass terms are generated in the Lagrangian (\ref{eq:4fermion})
beside the bare mass terms. 

\begin{equation}
L_{\psi}\rightarrow L_{\Psi}^{\mathrm{lin}}=-m_{+}\overline{\Psi_{D}^{+}}\Psi_{D}^{+}-m_{1}\overline{\Psi_{D}^{0}}\Psi_{D}^{0}-m_{2}\overline{\Psi}_{S}\Psi_{S}-m_{3}\left(\overline{\Psi^{0}}_{D}\Psi_{S}+\overline{\Psi}_{S}\Psi_{D}^{0}\right),\label{eq:fermion mass}\end{equation}
with%
\begin{figure}[h]
\begin{center}\includegraphics[%
  scale=0.45]{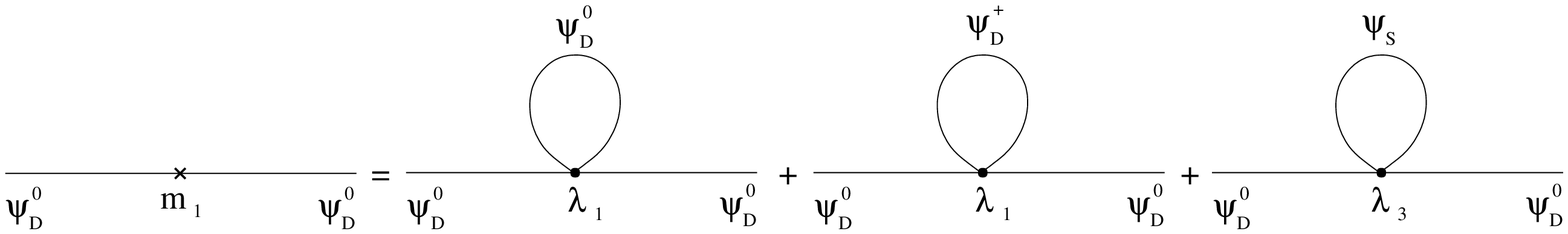}\end{center}

\begin{center}\textbf{Figure 1.} Feynman graphs for the gap equation
(\ref{eq:mtable}). Similar graphs corresponding to (\ref{eq:mtablp},\ref{eq:mtabl2})
with exchanged legs and lines.\end{center}
\vspace{0.3cm}

\begin{center}\includegraphics[%
  scale=0.24]{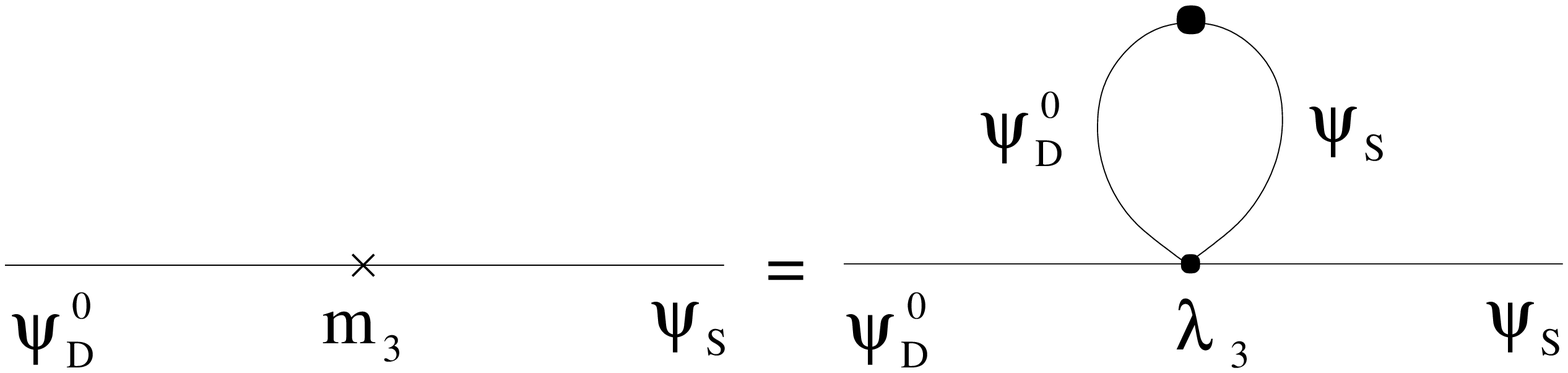}\end{center}

\begin{center}\textbf{Figure 2.} Feynman graphs for the gap equation
(\ref{eq:mtabl3}).\end{center}\vspace{0.3cm}

\end{figure}

\begin{eqnarray}
m_{+} & = & m_{0D}-6\lambda_{1}a_{+}-8\left(\lambda_{1}a_{1}+\lambda_{3}a_{2}\right)=m_{1}+2\lambda_{1}\left(a_{+}-a_{1}\right)\label{eq:mtablp}\\
m_{1} & = & m_{0D}-6\lambda_{1}a_{1}-8\left(\lambda_{1}a_{+}+\lambda_{3}a_{2}\right),\label{eq:mtable}\\
m_{2} & = & m_{0S}-6\lambda_{2}a_{2}-8\lambda_{3}\left(a_{1}+a_{+}\right),\label{eq:mtabl2}\\
m_{3} & = & 2\lambda_{3}a_{3}.\label{eq:mtabl3}\end{eqnarray}
If $m_{3}=0$ ($\lambda_{3}=0$ or $a_{3}=0$) then (\ref{eq:fermion mass})
is diagonal, the original gauge eigenstates are the physical fields,
the electroweak symmetry is not broken, $\lambda_{3}a_{3}$, the non-diagonal
condensate triggers the mixing and symmetry breaking. If $m_{3}\neq0$
(\ref{eq:fermion mass}) is diagonalized via unitary transformation
to get physical mass eigenstates 

\begin{eqnarray}
\Psi_{1} & = & \phantom{-}c\,\Psi_{D}^{0}+s\,\Psi_{S},\nonumber \\
\Psi_{2} & = & -s\,\Psi_{D}^{0}+c\,\Psi_{S},\label{eq:fermion mixing}\end{eqnarray}
where $c=\cos\phi$ and $s=\sin\phi$, $\phi$ is the mixing angle.
The masses of the physical fermions $\Psi_{1},\:\Psi_{2}$ are

\begin{equation}
2M_{1,2}=m_{1}+m_{2}\pm\frac{m_{1}-m_{2}}{\cos2\phi}.\label{eq:mphys}\end{equation}
The mixing angle is defined by \begin{equation}
2m_{3}=(m_{1}-m_{2})\tan2\phi.\label{eq:def phi}\end{equation}
Again we see, once $m_{3}=0$ the mixing angle vanishes (for $m_{1}\neq m_{2}$),
$M_{1}=m_{1}$ and $M_{2}=m_{2}$. The physical masses may be equal
($M_{1}=M_{2})$ only if $m_{1}=m_{2}$, the original neutral fermions
are degenerate in mass and then the mixing angle is meaningless from
the point of view of mass matrix diagonalization.

It follows that the physical eigenstates themselves form condensates
since\begin{eqnarray}
c^{2}\left\langle \overline{\Psi}_{1\alpha}\Psi_{1\beta}\right\rangle _{0}+s^{2}\left\langle \overline{\Psi}_{2\alpha}\Psi_{2\beta}\right\rangle _{0} & = & a_{1}\delta_{\alpha\beta},\nonumber \\
s^{2}\left\langle \overline{\Psi}_{1\alpha}\Psi_{1\beta}\right\rangle _{0}+c^{2}\left\langle \overline{\Psi}_{2\alpha}\Psi_{2\beta}\right\rangle _{0} & = & a_{2}\delta_{\alpha\beta},\label{eq:condphys}\\
cs\left\langle \overline{\Psi}_{1\alpha}\Psi_{1\beta}\right\rangle _{0}-cs\left\langle \overline{\Psi}_{2\alpha}\Psi_{2\beta}\right\rangle _{0} & = & a_{3}\delta_{\alpha\beta}.\nonumber \end{eqnarray}
There is no non-diagonal condesate as $\Psi_{1},\:\Psi_{2}$ are independent.
Combining the equations of (\ref{eq:condphys}) one finds

\begin{equation}
a_{3}=\frac{1}{2}\tan2\phi\left(a_{1}-a_{2}\right).\label{eq:a3rel}\end{equation}
For $a_{1}=a_{2}$, $a_{3}\neq0$ is not possible for $\cos2\phi\neq0$.
As is seen, (\ref{eq:a3rel}) is equivalent to $\left\langle \overline{\Psi}_{1\alpha}\Psi_{2\beta}\right\rangle _{0}=0.$
Comparing (\ref{eq:a3rel}) to (\ref{eq:def phi}) yields \begin{equation}
m_{1}-m_{2}=2\lambda_{3}\left(a_{1}-a_{2}\right).\label{eq:m1m2rel}\end{equation}
Using the equations (\ref{eq:mtablp}-\ref{eq:mtabl3}) we are lead
to a consistency conditions\begin{equation}
\left(\lambda_{3}-\lambda_{1}\right)\left(a_{1}+\frac{4}{3}a_{+}\right)=\left(\lambda_{3}-\lambda_{2}\right)a_{2},\label{eq:cons}\end{equation}
$\lambda_{1}\neq\lambda_{2}$ goes with $a_{1}+\frac{4}{3}a_{+}\neq a_{2}$.

The equations (\ref{eq:mtablp}-\ref{eq:mtabl3}) can be formulated
as gap equations in terms of the physical fields expressing both the
masses and the condensates with $\Psi_{1}$, $\Psi_{2}$ and $\Psi_{+}\equiv\Psi_{D}^{+}$.
Assuming vanishing original masses, $m_{0S}=0$, $m_{0D}=0$, the
complete set of gap equations are\begin{eqnarray}
c\cdot s\left(M_{1}-M_{2}\right) & = & 2\lambda_{3}\; c\cdot s\left(I_{1}-I_{2}\right),\label{eq:gap3}\\
c^{2}M_{1}+s^{2}M_{2} & = & -\lambda_{1}\left(6\left(c^{2}I_{1}+s^{2}I_{2}\right)+8I_{+}\right)-8\lambda_{3}\left(s^{2}I_{1}+c^{2}I_{2}\right),\label{eq:gap1}\\
s^{2}M_{1}+c^{2}M_{2} & = & -6\lambda_{2}\left(s^{2}I_{1}+c^{2}I_{2}\right)-8\lambda_{3}\left(c^{2}I_{1}+s^{2}I_{2}+I_{+}\right),\label{eq:gap2}\\
M_{+} & = & -\lambda_{1}\left(8\left(c^{2}I_{1}+s^{2}I_{2}\right)+6I_{+}\right)-8\lambda_{3}\left(s^{2}I_{1}+c^{2}I_{2}\right).\label{eq:gap+}\end{eqnarray}

The main task of the present work is to explore the structure of the
gap equations. There are four algebraic equations for four variables
$M_{1},\; M_{2},\; M_{+}$, $c^{2}=\cos^{2}\phi$. As in almost all
approximation $I_{i}\sim M_{i}$, (\ref{eq:gap3}-\ref{eq:gap+})
show gap equation characteristics, $M_{i}=0$ is always a symmetric
solution, which is stable for small $|\lambda_{i}|$. Increasing $|\lambda_{i}|$
also an energetically favoured \cite{klev} massive solution emerges
as in the original Nambu Jona-Lasinio model. Now we explore the parameter
space $\lambda_{i}$ to find acceptable phyical masses.

Let the condensates be approximated by free field propagators\begin{equation}
\left\langle \overline{\Psi}_{i\alpha}\Psi_{i\beta}\right\rangle =\frac{\delta_{\alpha\beta}}{4}I_{i}=-\frac{\delta_{\alpha\beta}}{8\pi^{2}}M_{i}\left(\Lambda^{2}-M_{i}^{2}\ln\left(1+\frac{\Lambda^{2}}{M_{i}^{2}}\right)\right),\quad i=1,2,+,\label{eq:free1}\end{equation}
where $M_{+}=m_{+}$. Here $\Lambda$ is a four-dimensional physical
cutoff, it sets the scale of the new physics responsible for the non-renormalizable
operators. From the point ov view of symmetry breaking $\Lambda$
can be chosen arbitrary large (below the GUT or Planck scale), but
higher $\Lambda$ implies stronger fine tuning of $\lambda_{3}$,
see (32), to keep the new fermion masses in the electroweak range.
To avoid fine tuning and allow reasonable fermion masses $\Lambda$
is expected to be a few TeV, typically around 3 TeV \cite{fcm}.

For the electroweak symmetry breaking the most important equation
is (\ref{eq:gap3}), it triggers mixing between the different representations
of the weak gauge group. Applying (\ref{eq:free1}) it reads

\begin{equation}
0=\left(M_{1}-M_{2}\right)c\cdot s\left(\frac{1}{\lambda_{3}}+\frac{\Lambda^{2}}{\pi^{2}}-\frac{M_{1}^{3}\ln\left(1+\frac{\Lambda^{2}}{M_{1}^{2}}\right)-M_{2}^{3}\ln\left(1+\frac{\Lambda^{2}}{M_{2}^{2}}\right)}{M_{1}-M_{2}}\right).\label{eq:gap3an}\end{equation}
(\ref{eq:gap3an}) always has a symmetric solution $\left(M_{1}-M_{2}\right)c\cdot s=0$,
implying $\sin2\phi=0$ for $M_{1}\neq M_{2}$ ,essentialy no mixing,
$M_{1}=M_{2}$ is discussed after (\ref{eq:def phi}). If $|\lambda_{3}|$
is greater than a critical value $\left|\lambda_{3}^{c}\right|=\frac{\pi^{2}}{\Lambda^{2}}$
there also exists a symmetry breaking solution ($M_{1}\neq M_{2})$,
which always has lower energy if the massive solution exists \cite{klev}.
Equation (\ref{eq:gap3an}) has a solution with moderate masses ($M_{1,2}<0.7\Lambda$
) if $\lambda_{3}$ is negative. In the small mass limit the parantheses
in (\ref{eq:gap3an}) simplifies to $\frac{1}{\lambda_{3}}+\frac{\Lambda^{2}}{\pi^{2}}-\left(M_{1}^{2}+M_{1}M_{2}+M_{2}^{2}\right)\left(\ln\left(\Lambda^{2}\right)-\ln\left(\tilde{M^{2}}\right)\right)$
where $\tilde{M}\simeq max(M_{1},M_{2})$. If $|\lambda_{3}|$ is
slightly larger than its critical value, then we generally get small
masses compared to $\Lambda$, $M_{1}^{2}+M_{1}M_{2}+M_{2}^{2}\ll\Lambda^{2}$.
The critical coupling agrees with the original Nambu-Jona Lasinio
value, only a factor of two coming from the definition in the Lagrangian
(\ref{eq:4fermion}), $\lambda_{3}$ also defined differently in \cite{fcm}.
If $\left|\lambda_{3}\right|<\left|\lambda_{3}^{c}\right|$ then the
parantheses does not vanish in (\ref{eq:gap3an}), the condensate
$a_{3}$ is not formed and $\left(M_{1}-M_{2}\right)c\cdot s=0$.
The physical solution is $c\cdot s=0$, there is no meaningful mixing,
$\Psi_{S},\,\Psi_{D}$ are the physical mass eigenstates, and the
electroweak symmetry is not broken.

\begin{figure}
\begin{center}\includegraphics[%
  scale=0.75]{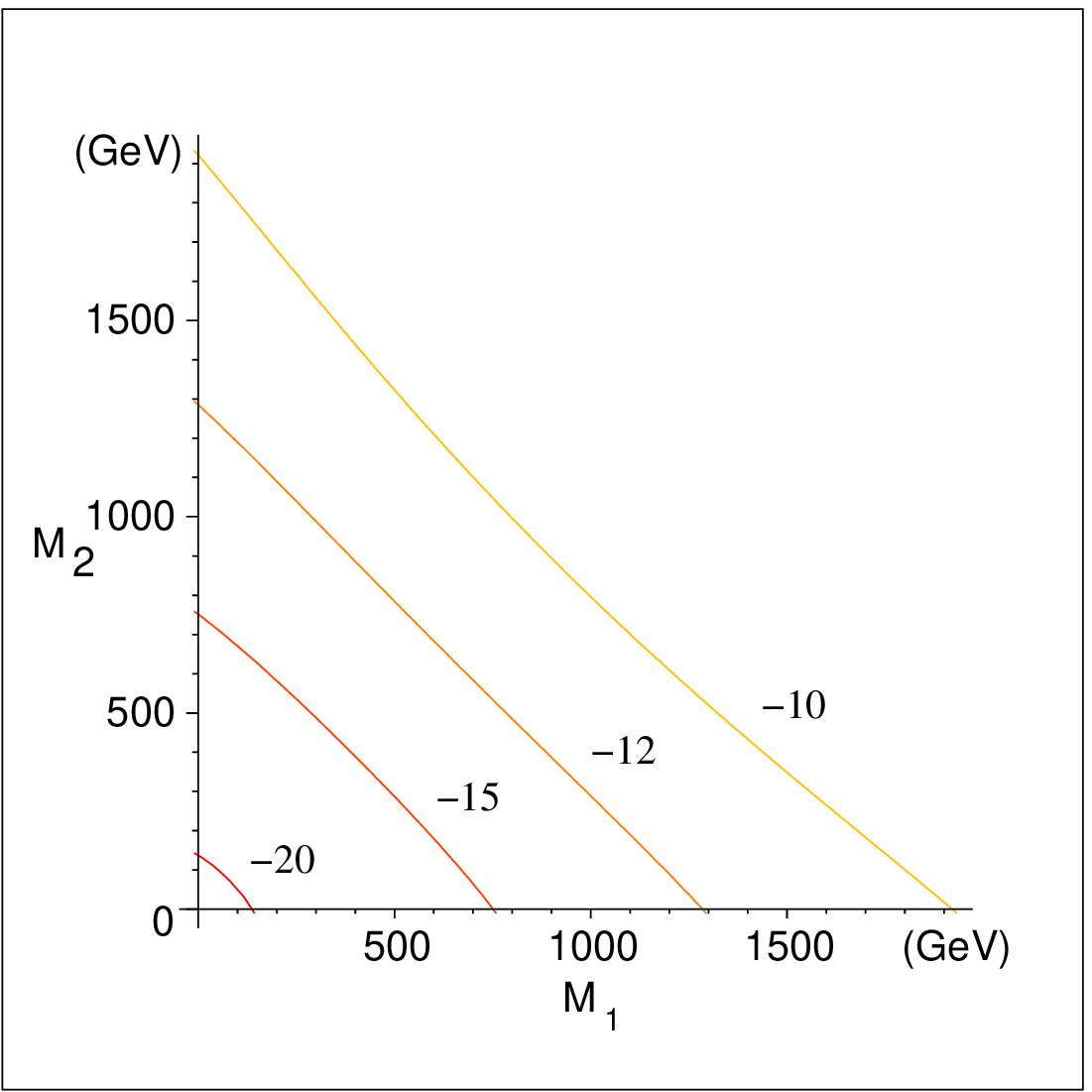}\end{center}

\begin{center}\textbf{Figure 3.} Constant $\lambda_{3}$ contours
in the $M_{1}$-$M_{2}$ plane for $\lambda_{3}=\left\{ -10,-12,-15,-20\right\} \cdot1/\Lambda^{2}$,
$\Lambda=3\;$TeV.\end{center}
\end{figure}

Despite the complicated structure of the non-linear equations (\ref{eq:gap3}-\ref{eq:gap+})
we get a relatively simple gap equation for $\lambda_{1}$, similar
to (\ref{eq:gap3an}), from (\ref{eq:mtablp}) $2\lambda_{1}\left(a_{1}-a_{+}\right)=m_{1}-m_{+}$.
In the physical fields we have\begin{equation}
M_{+}-c^{2}M_{1}-s^{2}M_{2}=2\lambda_{1}\left(I_{+}-c^{2}I_{1}-s^{2}I_{2}\right).\label{eq:l1an}\end{equation}
It includes four unknowns, therefore it cannot be analyzed directly.
We get a useful restriction solving (\ref{eq:gap3}) and (\ref{eq:gap1})
for $\lambda_{1}$ and substituting it to (\ref{eq:l1an}), relating
$M_{1},M_{2},M_{+}$ and $c^{2}$ independently of the $\lambda_{i}$'s.
Requiring that $0\leq c^{2}\leq1$ we get \begin{equation}
M_{1}\leq M_{+}\leq M_{2}.\label{eq:mpconstr}\end{equation}
As a result of the logaritmic terms in $I_{i}$, $M_{+}$ is nonlinear
in $c^{2}$, while $m_{1}=c^{2}M_{1}+s^{2}M_{2}$. We remark that
though (\ref{eq:gap3}) and (\ref{eq:l1an}) are very similar, for
moderate masses $\lambda_{3}$ is always negative, while $\lambda_{1}$
is positive (also $\lambda_{2}>0$). In the $c^{2}=1\;(0)$ limit
$M_{+}=M_{1}\;(M_{2})$ and there are cancelations in (\ref{eq:gap3}-\ref{eq:gap+}).
Turning back to the symmetric solution of (\ref{eq:gap3an}) the relation
(\ref{eq:mpconstr}) gives $M_{+}=M_{1}=M_{2}$ and the rest of the
gap equations set the common mass equal to zero unless the special
relation $6(\lambda_{3}-\lambda_{2})=8(\lambda_{3}-\lambda_{1})$
holds to provide cancellations.

To find the critical value for $\lambda_{1}$ and $\lambda_{2}$ we
considered the limit $M_{+}\rightarrow M_{2}=M$ and $M_{1}\rightarrow0$
then \begin{equation}
\lambda_{1}=\frac{1}{7}\frac{\pi^{2}}{\Lambda^{2}-M^{2}\ln\left(1+\frac{\Lambda^{2}}{M^{2}}\right)},\quad\lambda_{2}=\frac{4}{3}\frac{\pi^{2}}{\Lambda^{2}-M^{2}\ln\left(1+\frac{\Lambda^{2}}{M^{2}}\right)}.\label{eq:l1crit}\end{equation}
We get the same NJL type expression if we take the limit $M_{+}\rightarrow M_{1}=M$
and $M_{2}\rightarrow0$. (\ref{eq:l1crit}) provides massive solutions
if $\lambda_{1}\geq\frac{1}{7}\frac{\pi^{2}}{\Lambda^{2}}$ and $\lambda_{2}\geq\frac{4}{3}\frac{\pi^{2}}{\Lambda^{2}}$
. Numerical scans show that these are the minimal, critical values
for the couplings and can be approximated in special limits. Numerical
solutions are shown in Table 1. for cutoff $\Lambda=3$ TeV. The role
of $M_{1}$ and $M_{2}$ can be exchanged together with $c^{2}\leftrightarrow s^{2}$,
therefore we chose $M_{1}<M_{2}$ without the loss of generality.
As the cutoff is not too high, 3 TeV, there is no serious fine tuning
in the $\lambda_{i}$'s to find relatively small masses. 

To understand the signs and roughly the factors in $\lambda_{1,2}^{c}$
consider the limit $M_{1}\simeq M_{2}\simeq M_{+}\simeq M$. If $M\ll\Lambda$
then $\lambda_{3}\simeq\lambda_{3}^{c}=-\frac{\pi^{2}}{\Lambda^{2}}$,
though in the exact limit (\ref{eq:gap3}) becomes singular. We get
from (\ref{eq:gap3}-\ref{eq:gap+}) the relation $14\lambda_{1}=6\lambda_{2}+8\lambda_{3}$
and a single gap equation ( $I=I_{M}$ in (\ref{eq:free1}) )\begin{equation}
M=-\left(14\lambda_{1}+8\lambda_{3}\right)I.\label{eq:smallm}\end{equation}
Small mass solution requires $\tilde{\lambda}=14\lambda_{1}+8\lambda_{3}$
to be close to it's critical value $2\pi^{2}/\Lambda^{2}$ and provides
rough estimates $\lambda_{1}\sim\frac{5}{7}\frac{\pi^{2}}{\Lambda^{2}}$
and also $\lambda_{2}\sim3\frac{\pi^{2}}{\Lambda^{2}}$ to generate
small masses. Numerical solutions also provide general ($M_{+}$ not
close to $M_{1}$ or $M_{2}$) small masses for couplings close to
these values, see Table 1.

In the strongest small mass limit one neglects the logaritmic terms
in the condensates (\ref{eq:free1}), and equations (\ref{eq:gap3}-\ref{eq:gap+})
reduce to a linear homogeneous system of equations. Two conditions
emerge to find a nonzero solution for the physical masses \begin{eqnarray}
\left(1+2\lambda_{1}\frac{\Lambda^{2}}{\pi^{2}}\right)\left(M_{+}-c^{2}M_{1}-s^{2}M_{2}\right) & = & 0\label{eq:constr1}\\
\left(1-14\lambda_{1}\frac{\Lambda^{2}}{\pi^{2}}\right)\left(1-6\lambda_{2}\frac{\Lambda^{2}}{\pi^{2}}\right) & = & 128\left(\lambda_{3}\frac{\Lambda^{2}}{\pi^{2}}\right)^{2}\label{eq:constr2}\end{eqnarray}
Remember that $m_{+}=M_{+}$ and $m_{1}=c^{2}M_{1}+s^{2}M_{2}$, $m_{2}=s^{2}M_{1}+c^{2}M_{2}$.
In case $\lambda_{1}\frac{\Lambda^{2}}{\pi^{2}}=-\frac{1}{2}$ (\ref{eq:constr2})
becomes $\left(1-6\lambda_{2}\frac{\Lambda^{2}}{\pi^{2}}\right)=16\left(\lambda_{3}\frac{\Lambda^{2}}{\pi^{2}}\right)^{2}$
and there is a mass relation $m_{1}+m_{+}=\left(2\lambda_{3}\frac{\Lambda^{2}}{\pi^{2}}\right)m_{2}$.
If $\lambda_{1}\frac{\Lambda^{2}}{\pi^{2}}\neq-\frac{1}{2}$, we get
two relations for the masses, $m_{+}=m_{1}=c^{2}M_{1}+s^{2}M_{2}$
and $\frac{m_{1}}{m_{2}}=\frac{1-6\lambda_{2}\Lambda^{2}/\pi^{2}}{16\lambda_{3}\Lambda^{2}/\pi^{2}}=\frac{8\lambda_{3}\Lambda^{2}/\pi^{2}}{1-14\lambda_{1}\Lambda^{2}/\pi^{2}}$.

\begin{table}[th]
\begin{center}\begin{tabular}{|c|c|c|c|c|c|c|c|}
\hline 
$\lambda_{1}$$\left(\frac{\pi^{2}}{\Lambda^{2}}\right)$&
0.546&
0.740&
0.496&
0.380&
0.502&
0.468&
0.419\tabularnewline
\hline 
$\lambda_{2}$$\left(\frac{\pi^{2}}{\Lambda^{2}}\right)$&
2.540&
3.11 (!)&
2.403&
2.120&
2.457&
2.455&
2.451\tabularnewline
\hline 
$\lambda_{3}$$\left(\frac{\pi^{2}}{\Lambda^{2}}\right)$&
-1.031&
-1.041&
-1.042&
-1.070&
-1.083&
-1.178&
-1.330\tabularnewline
\hline
\hline 
$M_{1}$ (GeV)&
100&
148&
100&
100&
150&
200&
200\tabularnewline
\hline 
$M_{2}$ (GeV)&
150&
150&
200&
300&
300&
500&
800\tabularnewline
\hline 
$M_{+}$ (GeV)&
149&
149&
190&
290&
290&
490&
790\tabularnewline
\hline
\end{tabular}\end{center}

\caption{Solutions of the gap equations for the cutoff $\Lambda=$3 TeV, $\lambda_{i}$
are given in units of $\frac{\pi^{2}}{\Lambda^{2}}$. In the second
column $\lambda_{2}$ violates perturbative unitarity.}
\end{table}

We have calculated and presented briefly in \cite{fcm} the constraints
on the parameters of the model from perturbative unitarity \cite{unit,unitvcm}.
Consider the amplitudes of two particle $\left(\Psi_{D}^{(+)},\Psi_{D}^{(0)}\,\mathrm{or}\;\Psi_{S}\right)$
elastic scattering processes and impose $\left|\Re a_{0}\right|\leq1/2$
for the $J=0$ partial wave amplitudes. The contact graph gives the
dominant contribution, neglecting the fermion masses for the $\Psi_{D}^{(+)}\Psi_{D}^{(-)}$
scattering gives an upper bound on $\lambda_{1}$ coupling, $\left|\lambda_{1}\right|s\leq8\pi$,
where $s$ is the maximal center of mass energy $\left(M_{+}^{2}\ll s\leq\Lambda^{2}\right)$. 

We cannot always use the small mass limit, as the solution of the
gap equations provide higher $\lambda_{i}$'s for significantly higher
masses. Therefore we have calculated different helicity amplitudes
\cite{unitfermion} for non-vanishing masses. For $\Psi_{a}(1)\bar{\Psi}_{a}(2)\rightarrow\Psi_{a}(3)\bar{\Psi}_{a}(4)$,
($a=0,\, s,+$), $M=\lambda_{i}\left[\left(\bar{v}_{2}u_{1}\right)\left(\bar{u}_{3}v_{4}\right)-\left(\bar{u}_{3}u_{1}\right)\left(\bar{v}_{2}v_{4}\right)\right]$,
where $\lambda_{i=1,2,3}$ is the only relevant four-fermion coupling.
We consider $\Psi_{S}=s\Psi_{1}+c\Psi_{2}$ scattering as a linear
combination in the coupled $\Psi_{1},\,\Psi_{2}$ channels to employ
only $\lambda_{2}$ (and simiarly $\Psi_{D}^{(0)}$ to constrain $\lambda_{1}$).
The contributions of the $\gamma,\, Z$ exchange graphs are negligible
$\left({\cal O}\left(g^{2}\right)\ll8\pi\right)$ because of the extra
propagator. There are three different helicity channels, we give the
representative helicity amplitudes, these are maximal for the back
to back scattering ($\theta_{\mathrm{scattering}}^{\left\{ 13\right\} }=\pi$)
\begin{eqnarray}
M\left((+-)\rightarrow(+-)\right) & = & \lambda_{i}\left(s-4M_{i}^{2}\right),\label{eq:hel1}\\
M\left((++)\rightarrow(--)\right) & = & \lambda_{i}s,\label{eq:hel2}\\
M\left((+-)\rightarrow(-+)\right) & = & \lambda_{i}4M_{i}^{2}.\label{eq:hel3}\end{eqnarray}
For other scattering angles $\left|M\right|$ is smaller than in (\ref{eq:hel2}),
for example the maximum for $\theta=0$ is $\lambda_{i}4M_{i}^{2}$.
The mass dependent unitarity bound agrees with the first estimate\begin{equation}
\lambda_{i}s\leq8\pi,\label{eq:unit}\end{equation}
where (i=1,2,3) and $s\leq\Lambda^{2}$ is the center of mass energy.
The unitarity constraints are most stringent for $\lambda_{2}$, even
the equal small mass limit (\ref{eq:smallm}) would set $\lambda_{2}\simeq3\pi^{2}/\Lambda^{2}$
which is above the maximum allowed by unitarity $8\pi/\Lambda^{2}\simeq2.55\cdot\frac{\pi^{2}}{\Lambda^{2}}$.
As an example we show a non-physical nearly equal mass solution in
the second column of Table 1., which is not allowed by perturbative
unitarity. (\ref{eq:unit}) implies an absolute upper bound on the
smaller neutral mass, $M_{1}\leq230$ GeV for $\Lambda=3$TeV and
generally pushes up the charged mass close to $M_{2}.$ The $\lambda_{2}$
contours are drawn in the $M_{2}-M_{1}$plane for $M_{+}=M_{2}-10$GeV
in Figure 4. 

\begin{figure}
\begin{center}\includegraphics[%
  scale=0.75]{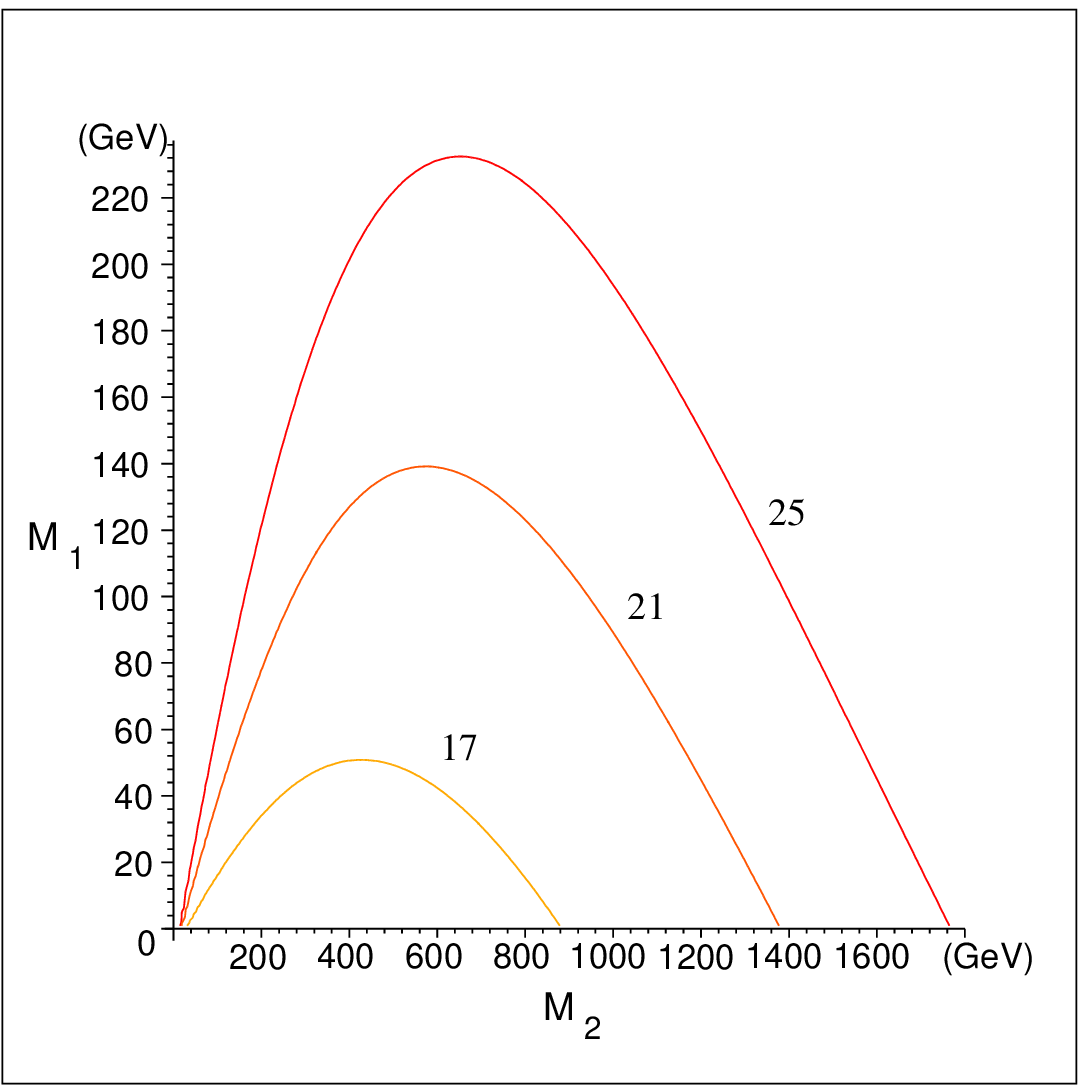}\end{center}

\begin{center}\textbf{Figure 4.} Constant $\lambda_{2}$ contours
in the $M_{2}$-$M_{1}$ plane for $\lambda_{2}=\left\{ 25,21,17\right\} 1/\Lambda^{2}$,
the allowed region is below the curves, $\Lambda=3\;$TeV.\end{center}
\end{figure}

In conclusion, we explored and solved the gap equations in an extension
of the recently proposed fermion condensate model of electroweak interactions
\cite{fcm}. The original fermions mix via the non-diagonal four-fermion
term in the non-renormalizable Lagrangian. The gap equations can be
formulated in the physical degrees of freedoms and we have given examples
of physical masses for various coupling constants. We determined the
critical couplings to define the region of massive solutions, $\lambda_{3}<-\pi^{2}/\Lambda^{2}$
is required to break the electroweak symmetry. With a few TeV cutoff
the couplings do not have to be fine tuned to imply masses of few
hundred GeV. In the spectrum three new fermions appear, between the
lightest and the heaviest neutral ones there is a charged particle.
Perturbative unitarity sets (via the coupling $\lambda_{2}$) the
mass of the charged fermion relatively close to the heavier neutral
one and the lightest fermion mass must be below an upper bound, $M_{1}\leq230$
GeV for $\Lambda=3$ TeV. The lightest new fermion is stable and a
good dark matter candidate as the new fermions interact only in pairs
and weakly with standard particles. The model can be tested soon at
LHC, numbers of pairs of new particles are expected with masses of
few hundred GeV, cross sections for electron-positron colliders are
presented in \cite{fcm}.

\subsubsection*{Acknowledgement}

The authors thank George Pócsik for valuable discussion and collaboration
on \cite{fcm}.

\end{document}